\documentclass[twocolumn,aps,nofootinbib,pra]{revtex4}

\usepackage{mathrsfs}
\usepackage{amsfonts}
\usepackage{amsmath}
\usepackage[colorlinks,citecolor=blue,linkcolor=blue]{hyperref}

\newcommand{\be}{\begin{equation}}
\newcommand{\ee}{\end{equation}}
\newcommand{\ba}{\begin{array}}
\newcommand{\ea}{\end{array}}
\newcommand{\bw}{\begin{widetext}}
\newcommand{\ew}{\end{widetext}}

\newcommand{\ra}{\rangle}
\newcommand{\la}{\langle}
\newcommand{\pp}{\partial}
\newcommand{\ov}{\overline}

\newcommand{\ww}{\widetilde}

\newcommand{\bk}{{\bf k}}
\newcommand{\bp}{{\bf p}}

\newcommand{\bq}{{\bf q}}

\newcommand{\cs}{\mathcal{S}}

\newcommand{\E}{{\cal E}}

\newcommand{\LL}{\mathcal{L}}
\newcommand{\N}{{\mathscr{N}}}

\newcommand{\PP}{{\mathcal{P}}}

\newcommand{\VV}{{\mathscr{V}}}

\newcommand{\tr}{{\rm Tr}}

\begin{document}

 \title{
 A quantum mechanism underlying the gauge symmetry in quantum electrodynamics
 }

\author{Wen-ge Wang\footnote{ Email address: wgwang@ustc.edu.cn}}
\affiliation{
 Department of Modern Physics, University of Science and Technology of China,
 Hefei 230026, China
 }

 \date{\today}

\begin{abstract}

 In this paper, a formulation, which is completely established on a quantum ground, is presented 
 for basic contents of quantum electrodynamics (QED).
 This is done by moving away, from the fundamental level, the assumption that 
 the spin space of bare photons should (effectively) possess the same properties 
 as those of  free photons observed experimentally. 
 Within this formulation,  bare photons with zero momentum can not be neglected
 when constructing the photon field;
 and an explicit expression for the related part of the photon field is derived.
 When a  local gauge transformation is performed on the electron field,
 this expression predicts a change that turns out to be
 equal to what the gauge symmetry requires for the gauge field.
 This gives an explicit mechanism, by which the photon field may change under gauge transformations in QED.
\end{abstract}

 \maketitle


\section{Introduction}\label{sect-intro}

 Quantum field theory (QFT) supplies a framework for establishing the most successful 
 model for elementary particles, i.e., the standard model (SM) \cite{Weinberg-book,Peskin,Itzy}.
 However, there are still some important problems unsolved within QFT.
 In this paper, we discuss some aspects of two of them.
 One problem is that the physical origin of gauge symmetry is still unclear. 
 As is well known, to a large extent, modern physics is established on two symmetries,
 the Lorentz symmetry and the gauge symmetry. 
 The physical origin of the Lorentz symmetry is quite clear, i.e., the physics should not depend on
 the frame of reference chosen. 
 In contrast, presently, no physical mechanism is known for the gauge symmetries employed in the SM.

 Another problem is that, although  elementary particles 
 are quantum particles in their nature, in the establishment of the SM, classical fields must be made use of 
 at the fundamental level.
 Specifically, in the canonical quantization formalism, quantum fields are introduced by quantizing classical fields;
 and in the path-integral quantization formalism integrals are computed for all realizations of classical fields. 
 Efforts have been seen in the direction of putting QFT completely on a quantum ground
 (see, e.g., Ref.\cite{Weinberg-book}), however, serious difficulties still lie in the way. 
 In this paper, we go further in this direction, by simplifying some of the strategies and methods
 used in Ref.\cite{Weinberg-book}. 
 Interestingly, with some progresses achieved in this direction, 
 a clue is seen for a partial solution to the first problem discussed above.

 Basically, Ref.\cite{Weinberg-book} suggests a route consisting of three steps:
 (i) establishing the quantum state space, (ii) constructing the quantum fields, and (iii) building the total Lagrangian. 
 A method of accomplishing the first step is discussed there,
 however, hard difficulties are met at the second step with quantum gauge fields.
 In order to overcome these difficulties, help is asked for 
 classical fields in a way similar to the ordinary treatments;
 that is, the quantum gauge fields are obtained by quantizing
 classical fields with certain gauge fixing.

 Examining the hard difficulties mentioned above, we observe that they are related to
 an assumption made at the fundamental level of the theory,
 which states that the spin spaces of bare bosons should (effectively) possess the same properties as 
 those of free bosons observed experimentally (see Sec.\ref{sect-clarify} for detailed discussions). 
 Since in fact there is no generic physical principle that may lie behind this assumption,
 it is of interest to study what the theory may look like,
 if this assumption is moved away from the fundamental level of the theory. 
 To study this possibility, in this paper, 
 we consider the simplest part of the SM, namely, the quantum electrodynamics (QED).
 We are to show that no serious difficulty is met with the
 three-step route suggested in Ref.\cite{Weinberg-book}, if this assumption is moved away.

 Moreover, within the framework of QED thus established, a clue is seen for a deeper
 understanding of gauge symmetry.
 In fact, within it, the photon field constructed in the usual way is not complete
 --- lacking the contribution from bare photons with zero momentum. 
 We found that the only reasonable way of constructing 
 the part of the field for these bare photons
 is to make use of the vacuum fluctuations induced by them.
 Going further, an explicit expression of this part of the field is derived.
 It predicts that, under a local $U(1)$ gauge transformation performed on the electron field, 
 this part of the photon field should undergo a change,
 which is equal to what is required by the gauge symmetry for the transformation of the total gauge field.

 The paper is organized as follows. 
 In Sec.\ref{sect-clarify}, we analyze some difficulties met in the route of establishing QFT discussed in
 Ref.~\cite{Weinberg-book}, then, propose a strategy of modifying it.
 In Secs.\ref{sect-ss-field}, following the proposed strategy, we formulate the basic part of QED.
 In Sec.\ref{sect-null-A-field}, we derive an explicit expression for the part of the photon field 
 that is related to bare photons with zero momentum
 and, then, show that it supplies a mechanism for understanding gauge transformations of the gauge field. 
 Finally, conclusions and discussions are given in Sec.\ref{sect-conclusion}.
 
\section{Preliminary analysis}\label{sect-clarify}

 In this section,  we analyze some difficulties met with the above-discussed 
 three-step route suggested in Ref.\cite{Weinberg-book} in the attempt of establishing QFT on a quantum ground.
 We also discuss some method for improvement.

 We begin with recalling some details of the first two steps given in Ref.\cite{Weinberg-book}.
 At the first step, the state space is established.
 It is spanned, as usual, by states that are obtained 
 from all possible direct products of single-particle states of all particle species.
 Here, a single-particle state possesses a momentum part and a spin part;
 the momentum part is just given by an ordinary momentum state, 
 while, properties of the spin part may be species-dependent. 
 In Ref.~\cite{Weinberg-book}, properties of the spin part are determined
 by a mathematical method of induced representations \cite{little-group},
 together with the well-known spin-statistics relationship \cite{Pauli40,SW64}
 and an assumption mentioned in Sec.\ref{sect-intro}. 
 \footnote{ 
 As an interesting feature of this treatment, the Dirac equation for fermionic spin states emerges
 as a requirement from symmetry considerations. }
 In what follows, we indicate this assumption by $A_{\rm bbs}$,  with ``bbs'' standing for ``bare boson spin''. 
\begin{itemize}
  \item $A_{\rm bbs}$. The spin spaces of bare bosons should possess the same properties as 
  those of the related free bosons observed experimentally, at least effectively so. 
\end{itemize}

 At the second step, quantum fields are constructed.
 To do this, creation and annihilation operators are introduced from the particles states
 gotten at the first step. 
 The quantum fields are written in the ordinary way as integrals of products of creation and annihilation operators
 and single-particle wave functions. 
 For fermions, the quantum fields can be directly constructed, without meeting any serious problem.

 However, hard difficulties are met with bosons subject to the assumption $A_{\rm bbs}$.
 In fact, on one hand, polarization vectors have four components, 
 on the other hand,  the angular momentum of experimentally-observed 
 $W^\pm$-boson (same for $Z^0$-boson) has only three independent eigenstates. 
 Even worse, no four-vector quantum field can be constructed for photons,
 if they are required to possess only two independent polarization states. 
 These difficulties are so hard that, in order to circumvent them, Ref.~\cite{Weinberg-book}
 goes back to the ordinary treatments to QFT, in which quantum fields
 are obtained by quantizing classical fields.

 Now, we give analysis.
 From the above discussions, it is seen that the emergence of the above-discussed hard difficulties 
 is related to the assumption $A_{\rm bbs}$.
 One may note that there is, in fact, no generic physical principle 
 that this assumption may be based on.
 On the contrary, from the conceptual angle, the assumed relationship looks unusual, 
 because the concept of bare particle lies at the fundamental level of the theory, while, 
 this is unnecessary for the concept of experimentally-observed free particle.

 One may see the above-mentioned point about concept clearly in an imagined model,
 which contains no divergence.
 In such a model, properties of ``bare particles'' are assumed at the fundamental level.
 Their states give a basis, in which  matrices of Hamiltonians are written.
 Meanwhile, stable states of ''free particles'' observed experimentally 
 are usually interpreted as eigenstates of some Hamiltonians.

 In a QFT such as the SM, the situation is much more complicated than the imagined model, 
 due to various divergences, particularly the ultraviolet divergence. 
 For the purpose of giving finite predictions for experimental results, 
 renormalization is needed to move out the divergences. 
 As is known, the assumption $A_{\rm bbs}$ is indeed useful
 in the computation of the scattering matrix,
 whose elements are for states of free particles observed experimentally. 
 In fact, in order to get finite predictions in a model suffering of the ultraviolet divergence, 
 it is always necessary to introduce some assumption like $A_{\rm bbs}$, as a working assumption,
 which imposes a link between spin spaces at the fundamental level of the theory
 and those of free particles observed experimentally.

 A relevant question is about the stage, at which such an assumption may be introduced. 
 One method is to do it at the fundamental level, like in ordinary formulations of QFT.
 This may give the theory a relatively good appearance,
 but, usually at the cost of imposing a quite complicated and involved
 structure at the fundamental level, even at a risk of blurring some actual structure of the theory.

 In order to avoid the above-mentioned flaw, one may consider an alternative strategy, 
 in which such an assumption 
 is put at some level higher than the fundamental level.
 In other words, such an assumption may be introduced, after the fundamental
 part of the theory is established.
 Partially, this strategy is similar that adopted in the indefinite-metric scheme of canonical quantization of the 
 classical electromagnetic field  \cite{Gupta}.
 \footnote{For a difference between our treatment and the indefinite-metric scheme,
 see discussions given in the paragraph below Eq.(\ref{Amu-QED-H}).}

 Summarizing the above discussions, we propose to consider the following route,
 as a modification to the three-step  route discussed in Ref.\cite{Weinberg-book}.
\begin{enumerate}
  \item[R1.] At the fundamental level.
  \begin{enumerate}
    \item Establishing the quantum state space,
    \item constructing the quantum fields,
    \item building the total Lagrangian. 
  \end{enumerate}
  \item[R2.] At a higher level.
\begin{itemize}  
 \item Either introducing an assumption like $A_{\rm bbs}$,
  \item or interpreting eigenstates of some effective Hamiltonians, if obtainable, as 
    states of free particles observed experimentally. 
\end{itemize}
\end{enumerate}
 In this paper, we discuss only at the fundamental level
 and, hence, we do not need to consider any assumption like $A_{\rm bbs}$.
  
\section{Basic contents of QEDoM}\label{sect-ss-field}

 In this section, we discuss basic contents of QED at a fundamental level,
 not including the assumption $A_{\rm bbs}$.
 We first give some preliminary discussions in Sec.\ref{sect-pre-dis}, 
 then, discuss the quantum state spaces and quantum fields in Sec.\ref{sect-e} and Sec.\ref{sect-A},
 from which the ordinary Lagrangian can be built (Sec.\ref{sect-Lag}). 

\subsection{Preliminary discussions}\label{sect-pre-dis}

 We first give some words for terminology. 
 Since bare elementary particles are quite different, in their nature, 
 from free particles that are observed experimentally, 
 it would be convenient to give them names with more explicit distinction.
 For this reason, we use the term \emph{mode} to refer to
 a bare elementary particle that is considered at the fundamental level of QED;
 meanwhile, free particles observed experimentally are just called free particles. 
 (A further reason of making this distinction will be discussed in Sec.\ref{sect-conclusion}.)
 Specifically, instead of bare electron, bare positron, and bare photon, we say $e$-mode, $\ov e$-mode 
 (called anti $e$-mode), and $A$-mode. 
 Moreover, with the assumption $A_{\rm bbs}$ moved away from the fundamental level, 
 the formulation of the fundamental part of QED to be given below is not exactly the same as ordinary ones.
 We use ``\emph{QED of modes}'', in short, QEDoM, to refer to the formulation of QED to be given below.

 Like in the ordinary treatment,  
 the state of an arbitrary mode is assumed to possess two parts, a momentum part and a spin part.
 We use $\E_M$ to denote the state space of a single mode $M$, with $M=e, \ov e, A$,
 and, as usual, assume that it has the following form,
  \begin{gather}\label{EP}
  \E_M = \bigoplus_{\bp} |\bp\ra \otimes \cs_M,
  \end{gather}
 where $|\bp\ra$ indicate the ordinary momentum states 
 and $\cs_M$ is either a representation space of the Lorentz group or a subspace of it.
 The momentum states satisfy an ordinary normalization condition, i.e.,
\begin{gather}\label{<bp|bq>}
  \la \bq |\bp \ra   = p^0 \delta^3(\bp-\bq),
\end{gather}
 where $p^0 = \sqrt{|\bp|^2 + m_0^2}$, with $m_0$ indicating the mass of $e$-mode.
 The spin space $\cs_M$  may be a function of the momentum $\bp$.

 In Ref.~\cite{Weinberg-book}, to satisfy the assumption $A_{\rm bbs}$,
 properties of the spaces $\cs_M$ are determined by the method of induced representations,
 which makes use of representations of the so-called little group \cite{little-group}.
 But, in QEDoM, with the assumption $A_{\rm bbs}$ moved away from the fundamental level, 
 there is no need of using the method of little group to determine properties of $\cs_M$. 
 In fact, the physical reason of employing the method of little group is far from being apparent. 
 Instead, to determine properties of $\cs_M$, we are to employ 
 an alternative method that was discussed in Ref.\cite{pra16-commu};
 this method is based on the physical requirement that predictions for experimental results must 
 have definite and real values.

\subsection{States and fields of $e$-mode and $\ov e$-mode}\label{sect-e}

 In this section, we first discuss the state space $\E_e$ for a single $e$-mode, 
 then, discuss $\E_{\ov e}$ for an $\ov e$-mode.
 Finally, we construct the corresponding quantum fields.

 We assume that the spin space $S_e$ should satisfy two requirements:
 (i) being a subspace of the four-dimensional Dirac-spinor space
 and (ii) being a Hilbert space. 
 As is well known, a Hilbert space is a complete and complex linear space equipped with an inner product.
 For two arbitrary vectors $|\psi\ra$ and $|\phi\ra$, their inner product,
 denoted  by $(|\psi\ra, |\phi\ra)$, by definition should possess the following basic properties,
\begin{subequations}\label{Hs}
\begin{eqnarray}
 & &  (|\psi\ra,a|\phi_1\ra+b|\phi_2\ra) = a(|\psi\ra,|\phi_1\ra) + b (|\psi\ra ,|\phi_2\ra),
 \ \ \ \ \label{Hs-a}
 \\  & & (|\psi\ra , |\phi\ra) = (|\phi\ra, |\psi\ra)^*, \label{Hs-b}
  \\  & &  (|\psi\ra , |\psi\ra) \ge 0 \quad \text{with $(|\psi\ra , |\psi\ra) = 0$ iff $|\psi\ra =0$}, \label{Hs-c}
\end{eqnarray}
\end{subequations}
 where $a$ and $b$ are arbitrary complex numbers.
 A physical idea  underlying Eqs.(\ref{Hs-b})-(\ref{Hs-c}) is to ensure that the theory 
 may give predictions that take real and definite values.

 Spaces that satisfy the two requirements mentioned above 
 have been discussed in Ref.\cite{pra16-commu} for a fermion with a nonzero mass.
 According to results given there,  the simplest and most natural choice of $\cs_e$
 is that it is spanned by the two ordinarily-used Dirac spinors $U_{r}(\bp)$ of $r=0,1$.
 \footnote{In this treatment, the Dirac equation for spin states appears as the simplest and most natural
 condition, under which a subspace of the four-dimensional Dirac-spinor space is a Hilbert space. }
 These $U$-spinors satisfy the following  normalization condition,
\begin{gather}\label{<Ur|Us>-sp}
 U_{r}^{\dag}(\bp) \gamma^0 U_s(\bp) =2m_0 \delta_{rs}.
\end{gather}
 Thus, a basis of $\E_e$, denoted by $|e_{\bp r}\ra$, is written as
 \footnote{More consistently, one may write the spinors $U_{r}(\bp)$ in the abstract notation of 
 Dirac's ket as discussed in Ref.~\cite{pra16-commu}, such that $|e_{\bp r}\ra=  |U_{r}(\bp)\ra |\bp\ra$.}
\begin{gather}\label{|b>}
   |e_{\bp r}\ra=  U_{r}(\bp) |\bp\ra .
\end{gather}
 The label $r$ is raised by $\delta^{rs}$ and lowered by $\delta_{rs}$. 
 The bra of $|e_{\bp r}\ra$ is defined in the ordinary way, that is, 
 $\la e_{\bp r}| = \la \bp | U^\dag_{r}(\bp) \gamma^0$. 
 As a result, 
\begin{gather}\label{}
 \la e_{\bp r}| e_{\bq s} \ra = 2m_0 p^0 \delta^3(\bp-\bq) \delta_{rs} . 
\end{gather}

 Next, the $\ov e$-mode can be treated in a similar way. 
 The space $\cs_{\ov e}(\bp)$ is spanned by the two ordinarily-used Dirac spinors $V_{r}(\bp)$ of $r=0,1$,
 which satisfy the following  normalization condition,
\begin{gather}\label{<Vr|Vs>-sp}
 V_{r}^{\dag}(\bp) \gamma^0 V_s(\bp) = -2m_0 \delta_{rs}.
\end{gather}
 Due to the minus sign on the right-hand side (rhs) of Eq.(\ref{<Vr|Vs>-sp}),
 the space $\cs_{\ov e}(\bp)$ is not exactly a Hilbert space.
 But, since this minus sign appears for both of the two spinors $V_{r}(\bp)$, 
 practically, its effects can be easily moved away, 
 without affecting the physical idea of obtaining finite and real predictions. 
 Then, the state of a single $\ov e$-mode with a momentum $\bp$ is written as
\begin{gather} \label{|d>}
 |\ov e_{\bp r}\ra = V_{r}(\bp) |\bp\ra,
\end{gather}
 and the corresponding bra is written as $\la \ov e_{\bp r}| = \la \bp | V^\dag_{r}(\bp) \gamma^0$.

 Making use of direct products of the single-mode states discussed above, it is easy to construct the 
 total state spaces for $e$-mode and $\ov e$-mode. 
 Then, creation and annihilation operators for these two modes can be introduced by an ordinary method 
 \cite{Weinberg-book}.
 Obeying the spin-statistics relationship,
 they satisfy the following well-known anticommutation relations,
\begin{subequations} \label{bd-bddag=0}
\begin{gather}
  \{ b^{r\dag }(\bp) ,  b^{s\dag }(\bq ) \} =0,
 \\  \{ d^{r\dag }(\bp) ,  d^{s\dag }(\bq ) \} =0,
 \\ \{ b^{r\dag}(\bp) , d^{s\dag}(\bq) \} =0,
 \\  \{ b^r(\bp) ,  b^{s\dag }(\bq ) \} =p^0 \delta^{rs} \delta^3(\bp-\bq), \label{b-bdag=0}
 \\  \{ d^r(\bp) ,  d^{s\dag }(\bq ) \} =p^0 \delta^{rs} \delta^3(\bp-\bq). \label{d-ddag=0}
\end{gather}
\end{subequations}
 Note that there is no minus sign on the rhs of Eq.(\ref{d-ddag=0}).
 For single-mode states, one writes
\begin{subequations} \label{bd-|>}
\begin{gather}\label{}
 |e_{\bp r}\ra = b^{r\dag }(\bp)|0\ra, \quad \la e_{\bp r}| = \la 0| b^{r}(\bp),
 \\ |\ov e_{\bp r}\ra = d^{r\dag }(\bp)|0\ra, \quad \la \ov e_{\bp r}| = \la 0| d^{r}(\bp),
\end{gather}
\end{subequations}
 where $|0\ra$ indicates the vacuum state.

 Finally, we construct quantum fields for $e$-mode and $\ov e$-mode.
 A standard construction is given by  \cite{Weinberg-book}
\begin{subequations}\label{psi-psi+-QED}
\begin{gather}\label{psi-QED}
  \psi(x) = \int d\ww p \left(  b^r(\bp) U_{r}(\bp) e^{-ipx}
  + d^{r\dag}(\bp) V_{r}(\bp) e^{ipx} \right),
\\ \psi^\dag(x) = \int d\ww p \left(  b^{r\dag}(\bp)
 U_r^{\dag }(\bp) e^{ipx} +d^r(\bp) V_r^{\dag }(\bp)e^{-ipx} \right), \label{psi+-QED}
\end{gather}
\end{subequations}
 where $d\ww p = \frac{1}{p^0} d^3p$.
 \footnote{ In the literature, the factor $\frac{1}{p^0}$ in $d\ww p$ is sometimes written as
 $\frac{1}{\sqrt{p^0}}$.
 Here, we write this form of $d\ww p$, because it is Lorentz-invariant.
 Consistently, the anti-commutation relation for creation and annihilation operators
 has a factor $p^0$ [see Eq.(\ref{bd-bddag=0})].
 Some constant prefactor may be multiplied to the field $\psi$, which is not written explicitly for brevity.}
 Here and hereafter, by convention, double appearance of a same index, one in an upper position and
 the other in a lower position, implies summation over the index.
 \footnote{In this paper, repeated labels in a same type of  position do not imply summation. }
 Like in the standard treatment, 
 one assumes that the Lagrangian density for free $e$-mode and free $\ov e$-mode, 
 denoted by $\LL^0_{e\ov e}$, has the form of
\begin{gather}\label{L0-ep}
 \LL^0_{e\ov e} = \psi^\dag \gamma^0 (i \gamma^\mu \pp_\mu -m_0)\psi,
\end{gather}
 where $\pp_\mu \equiv \pp / \pp x^\mu$.

\subsection{States and fields of $A$-mode}\label{sect-A}

 In this section, we discuss the state space $\E_A$ for a single $A$-mode, 
 then, discuss  the corresponding quantum field. 
 
 The spin space $\cs_A$ is assumed to be a four-component vector space, denoted by $\VV$,
 which possesses a metric $g_{\mu\nu}$ with diagonal elements $[g_{\mu\mu}]= [1,-1,-1,-1]$
 and offdiagonal elements zero.
 Clearly, it is not a Hilbert space.
 An often-used basis of $\VV$ is given by the polarization vectors $\varepsilon_{\lambda}^\mu(\bk) $,
 corresponding to a given momentum $\bk$;
 they satisfy $\varepsilon^{*}_{\lambda,\mu}(\bk) \varepsilon_{\lambda'}^\mu(\bk) = g_{\lambda \lambda'}$.
 Then, the state space $\E_A$ for one $A$-mode is spanned by the following states,
\begin{gather}\label{}
 |A_{\bk \lambda}\ra = \varepsilon_{\lambda}^\mu(\bk) |\bk\ra .
\end{gather}
 The bra of $|A_{\bk \lambda}\ra$ is defined in the ordinary way, that is, 
\begin{gather}\label{}
  \la A_{\bk \lambda}|= \la \bk| \varepsilon_{\lambda \mu}^{ *}(\bk).
\end{gather}
 This gives that
\begin{gather}\label{<A|A>}
 \la A_{\bk \lambda}|A_{\bk' \lambda'}\ra = g_{\lambda \lambda'} k^0  \delta^{3}(\bk-\bk'),
\end{gather}
 where $k^0 = |\bk |$.

 Making use of the above-discussed states of single $A$-mode, 
 one may easily construct the total state space for $A$-modes.
 One may also introduce creation and annihilation operators, denoted by $a_{\lambda}^{\dag}(\bk)$
 and $a_{\lambda}(\bk)$, respectively, which satisfy the following commutation relations,
\begin{subequations}\label{aa-adag}
\begin{gather}
  [ a^\dag_\lambda(\bk) , a_{\lambda'}^{\dag}(\bk') ] =0, \label{aa}
\\  [ a_\lambda(\bk) , a_{\lambda'}^{\dag}(\bk') ] = g_{\lambda \lambda'} k^0  \delta^{3}(\bk-\bk'). \label{aa-dga}
\end{gather}
\end{subequations}
 Without the need of obeying the assumption $A_{\rm bbs}$, 
 the field for the $A$-mode, denoted by $A_\mu(x)$, can be directly constructed, 
\begin{gather}
 A_\mu(x) = \int d\ww k  a_{\lambda}(\bk) \varepsilon^{\lambda}_\mu(\bk) e^{-ikx}
  + a^{\dag}_\lambda(\bk) \varepsilon^{\lambda*}_\mu(\bk) e^{ikx}, \label{Amu-QED-H}
\end{gather}
 where $d\ww k = \frac{1}{k^0} d^3k$.

 One meets a normalization problem in the state space $\E_A$;
 that is, states in it can not be normalized in the ordinary way by making use of
 the scalar products given in Eq.(\ref{<A|A>}), due to the inhomogeneous sign of $g_{\lambda\lambda}$. 
 In fact, a similar problem is met in the  indefinite-metric scheme of canonical quantization of the 
 classical electromagnetic field  \cite{Gupta};
 there, the problem is solved by imposing an auxiliary condition 
 related to a specific gauge, which effectively imposes a restriction to the physical state space.
 But, this method  is not applicable here, because  at the fundamental level of the theory 
 we intend to formulate QEDoM in a gauge-independent way.

 We solve the above-discussed problem by a method discussed in Ref.\cite{pra16-commu}.
 As mentioned previously, the physical idea, which underlies 
 Eqs.(\ref{Hs-b})-(\ref{Hs-c}) for an inner product in a Hilbert space, 
 is to give real and definite values for predictions (for measurement results).
 As discussed in Ref.\cite{pra16-commu}, for the purpose of achieving this goal, 
 these two equations are in fact too restrictive
 and one may, instead, consider \emph{a generalized inner product}.
 For two arbitrary vectors $|\psi\ra$ and $|\phi\ra$, their generalized inner product
 $(|\psi\ra, |\phi\ra)$ possesses the following properties,
\begin{subequations}\label{Hs2}
\begin{eqnarray}
 & &  (|\psi\ra,a|\phi_1\ra+b|\phi_2\ra) = a(|\psi\ra,|\phi_1\ra) + b (|\psi\ra ,|\phi_2\ra),
 \ \ \ \ \label{Hs-a2}
 \\  & & (|\psi\ra ,|\phi\ra)^* = \beta (|\phi\ra, |\psi\ra), \label{Hs-b2}
  \\  & &  (|\psi\ra , \PP |\psi\ra) \ge 0 \quad \text{with
  $(|\psi\ra ,\PP |\psi\ra) = 0$ iff $|\psi\ra =0$},\hspace{0.8cm}
  \label{Hs-c2}
\end{eqnarray}
\end{subequations}
 where $\beta$ is a parameter and $\PP$ is a Lorentz-invariant operator.
 Clearly, at $\beta =1$ and $\PP = I$ being the identity operator,
 the generalized inner product reduces to the ordinary inner product.

 In an explicit construction of a generalized inner product for the $A$-mode, we consider the simplest choice
 of $\beta$, i.e., $\beta=1$.
 Let us focus on the space $\E_A$, because generalization of the discussions to be given below
 to the total state space for $A$-mode is straightforward. 
 A generic vector  in the space $\E_A$ is written as $|\psi\ra = \int d\ww k  c_{\bk}^{ \lambda}|A_{\bk \lambda}\ra$, 
 with $c$-number coefficients $c_{\bk}^{ \lambda}$.
 The corresponding bra is given by
\begin{gather}\label{}
 \la \psi| = \int d\ww k  \la A_{\bk \lambda}| c^{\lambda *}_{\bk}.
\end{gather}
 We define the symbol $(|\psi\ra ,|\phi\ra)$ in Eq.(\ref{Hs2}) in the ordinary way, that is, 
\begin{gather}\label{inner-defin}
 (|\psi\ra ,|\phi\ra) = \la \psi|\phi\ra.
\end{gather}

 An operators $\PP$ that satisfies Eq.(\ref{Hs-c2}) may be constructed by a method used in Ref.\cite{pra16-commu},
 that is, 
\begin{gather}\label{PP}
 \PP = \sum_\lambda \int d\ww k  |A_{\bk \lambda}\ra \la A_{\bk \lambda}|.
\end{gather}
 Since both $d\ww k$ and the label $\lambda$ are Lorentz invariant, this operator $\PP$ is Lorentz invariant. 
 Making use of Eq.(\ref{<A|A>}), it is straightforward to verify that
\begin{gather}\label{<A|P|A>}
 \la A_{\bk_1 \lambda_1}|\PP|A_{\bk_2 \lambda_2}\ra 
=  \delta_{\lambda_1 \lambda_2} k^0_1  \delta^{3}(\bk_1-\bk_2).
\end{gather}
 Then, it is easy to verify that Eq.(\ref{Hs-c2}) is satisfied with the operator $\PP$ in Eq.(\ref{PP}) 
 and, hence, the space $\E_A$ possesses a generalized inner product.

 The generalized inner product defined with the above operator $\PP$ is Lorentz invariant. 
 To see this point, let us consider another arbitrary vector $|\phi\ra$, 
 expanded as $|\phi\ra = \int d\ww k  C_{\bk}^{ \lambda}|A_{\bk \lambda}\ra$. 
 Making use of Eqs.(\ref{inner-defin})-(\ref{<A|P|A>}), 
 it is straightforward to find that 
\begin{gather}\label{(psi,PP-phi)}
 (| \psi\ra , \PP |\phi\ra) = \sum_\lambda \int d\ww k    c^{\lambda *}_{\bk }  C_{\bk}^{ \lambda}.
\end{gather}
 One notes that, under a Lorentz transformation, the momentum label $\bk$ of a coefficient, say,
 of $c_{\bk}^{ \lambda}$, changes, while, the value of $c_{\bk}^{ \lambda}$ does not change. 
 Then, from the rhs of Eq.(\ref{(psi,PP-phi)}), 
 one sees that the value of $(| \psi\ra , \PP |\phi\ra)$ should be Lorentz invariant. 
 Based on the above discussions, the vector $|\psi\ra$ may be normalized in the following way, 
\begin{gather}\label{}
 \sum_\lambda  \int d\ww k  |c_{\bk}^{ \lambda}|^2 =1. 
\end{gather}

\subsection{The interaction Lagrangian for $\psi(x)$ and $A_\mu(x)$}\label{sect-Lag}

 The interaction between the fields constructed in the previous sections can be introduced in the standard way,
 as briefly discussed below, 
 by assuming the invariance of the Lagrangian density under local $U(1)$ gauge transformations of
 the $e$-$\ov e$-mode fields. 
 We use tilde to indicate results of gauge transformations.

 Local $U(1)$ gauge transformations of the $e$-$\ov e$-mode fields take the following form,
\begin{subequations}\label{psi-e-gt-2}
\begin{gather}\label{psi-e-gt}
 \psi(x) \to \ww \psi(x) = e^{-i\theta(x)} \psi(x),
 \\ \psi^\dag(x) \to \ww \psi^\dag(x) =  e^{i\theta(x)} \psi^\dag (x).
\end{gather}
\end{subequations}
 Clearly, the Lagrangian density $\LL^0_{e\ov e}$ for the $e$-$\ov e$-mode field given in Eq.(\ref{L0-ep})
 is not invariant under the above transformations. 
 To keep the total Lagrangian invariant, as is well known, 
 the $A$-mode field should undergo the following transformation, 
\begin{gather}\label{Amu-gt-QED}
 A_{\mu}(x) \to \ww A_{\mu}(x)  = A_{\mu}(x) - \frac 1e \pp_\mu \theta(x).
\end{gather}
 The total Lagrangian density denoted by $\LL_{\rm QED}$, which has the form of
\begin{gather}\label{L-QED}
 \LL_{\rm QED} = \psi^\dag \gamma^0 (i \gamma^\mu D_\mu -m_0)\psi - \frac{1}{4} (F_{\mu\nu})^2,
\end{gather}
 is invariant under the above gauge transformations. 
 Here, $D_\mu$ indicates the covariant derivative,
\begin{gather}\label{Dmu-QED}
 D_\mu = \pp_\mu - i eA_{\mu}(x),
\end{gather}
 and
\begin{gather}\label{Fmn}
 F_{\mu\nu} = \pp_\mu A_\nu - \pp_\nu A_\mu.
\end{gather}
 The interaction Lagrangian density is  written as
\begin{gather}\label{Lint-QED}
 \LL_{\rm int}(x) =  e\psi^\dag(x) \gamma^0 \gamma^\mu  \psi(x) A_{\mu}(x).
\end{gather}

\section{The field of $A$-mode with zero momentum }\label{sect-null-A-field}

 In fact, the rhs of Eq.(\ref{Amu-QED-H}) is not complete in the construction of the quantum field for $A$-mode.
 This is because the field should include all states that lie in the space $\E_A$,
 while, the rhs of Eq.(\ref{Amu-QED-H}) does not include the state of $A$-mode with zero momentum.
 In fact, the polarization vectors $\varepsilon^{\lambda}_\mu(\bk)$ can not be defined
 for an $A$-mode with $k^\mu =0$.

 In this section, we discuss the field of $A$-mode with zero momentum.
 For brevity,  we call $A$-modes with zero momentum \emph{null $A$-modes}
 and use $A^{\rm np}_\mu(x)$ to indicate  the field for null $A$-modes. 
 Specifically, in Sec.\ref{sect-gen-null-A}, we discuss some properties that the null-$A$-mode field should possess,
 then, in Sec.\ref{sect-explict-null-A}, we discuss explicit expressions for the null-$A$-mode field.
 Finally, in Sec.\ref{sect-gauge-null-A}, we discuss changes of this field under
 local $U(1)$ gauge transformations of the $e$-$\ov e$-field. 

\subsection{Generic properties of the null-$A$-mode field}\label{sect-gen-null-A}

 It is natural to assume that the Lagrangian density for
 the interaction between the null-$A$-mode field and the $e$-$\ov e$-mode field,
 denoted by $\LL_{\rm int}^{\rm np}(x)$, has the same form as the ordinary one in Eq.(\ref{Lint-QED}), i.e., 
\begin{gather}\label{Lint0-QED}
 \LL_{\rm int}^{\rm np}(x) = e\psi^\dag(x) \gamma^0 \gamma^\mu  \psi(x) A^{\rm np}_\mu(x).
\end{gather}
 This interaction Lagrangian implies that null $A$-modes may generate quantum 
 fluctuations as virtual pairs, each consisting of an $e$-mode and an $\ov e$-mode.

 Since a null $A$-mode possesses zero energy,
 there is no physical reason to assume that it may possess  a nonzero angular momentum. 
 Hence, we assume that a null $A$-mode possesses no intrinsic degree of freedom.
 As  a consequence, a null $A$-mode has a one-dimensional state space.
 We use $a^{\rm np}$ and $a^{\rm np \dag}$ to indicate the annihilation
 operator and creation operator for null $A$-mode, respectively.
 Formally, one may write the null-$A$-mode field as
\begin{gather}
 A^{\rm np}_\mu(x) = \N_0 \Big( a^{\rm np} K_{\mu}(x)
  +  a^{\rm np \dag} K^{*}_{\mu}(x) \Big), \label{Amu-k0}
\end{gather}
 where  $\N_0$ is a normalization factor and $K_{\mu}(x)$ indicates some
 c-number four-component vector field.

 Due to zero energy,  there is in fact no restriction to the number of null $A$-modes that may exist.
 A reasonable assumption is that there may exist an infinite number of null $A$-modes.
 We use $|\infty^\text{np}\ra$ to indicate the normalized state of these null $A$-modes.
 The direct product of this state and the vacuum state $|0\ra$ is denoted by $|0^{\rm np}\ra$,
\begin{gather}\label{}
 |0^{\rm np}\ra \equiv |0\ra \otimes |\infty^\text{np}\ra.
\end{gather}
 The infinite number of null $A$-modes requires that the two operators $a^{\rm np}$ and $a^{\rm np \dag}$
 can not obey a commutation relation like that in Eq.(\ref{aa-dga}).
 In fact, if one required that $[a^{\rm np}, a^{\rm np \dag}] = 1$, then,
 the vector $a^{\rm np}|\infty^\text{np}\ra$ would be unnormalizable.

 Since adding/subtracting one null $A$-mode to/from the state $|\infty^\text{np}\ra$ should bring no change to it,
 the simplest assumption about the actions of $a^{\rm np}$ and $a^{\rm np \dag}$ is that
\begin{gather}\label{a0-a0dag-def}
 a^{\rm np} |\infty^\text{np}\ra = a^{\rm np \dag} |\infty^\text{np}\ra = |\infty^\text{np}\ra ,
\end{gather}
 which is the only specific requirement that is imposed here
 on the two operators $a^{\rm np}$ and $a^{\rm np \dag}$.
 This assumption requires realness of $K_{\mu}$, i.e.,
 $ K^*_{\mu}=K_{\mu}$.
 Note that the two operators $a^{\rm np}$ and $a^{\rm np \dag}$ should be commutable with all other
 creation and annihilation operators.
 Making use of this fact and Eqs.(\ref{Amu-k0})-(\ref{a0-a0dag-def}), one finds that
\begin{gather}
 A^{\rm np}_\mu(x) G |0^{\rm np}\ra = 2 \N_0 K_{\mu}(x) G |0^{\rm np}\ra , \label{Amu-k0-0np}
\end{gather}
 where $G$ represents an arbitrary function of operators that do not include $a^{\rm np}$ and $a^{\rm np \dag}$.
 Hence,  the null-$A$-mode field $A^{\rm np}_\mu(x)$ is effectively a $c$-number field
 and can be written in the form of $A^{\rm np}_\mu(x) = 2 \N_0 K_{\mu}(x)$.

\subsection{ Explicit expressions of the null-$A$-mode field}\label{sect-explict-null-A}

 In order to find an explicit expression for $K_{\mu}(x)$, one faces the following obstacle:
 Null $A$-modes by themselves possess no property that can be used to introduce a four-component vector.
 To solve this problem, the only conceivable method seems to
 make use of some effect of the quantum fluctuations,
 which are induced by null $A$-modes according to the interaction Lagrangian in Eq.(\ref{Lint0-QED}). 
 As discussed above, the fluctuations take the form of emergence and vanishing of virtual $e$-$\ov e$-mode pairs.

 Based on discussions given above (in this and the previous sections), 
 we propose that the null-$A$-mode field $A^{\rm np}_\mu(x)$ should possess the following properties.
\begin{enumerate}
  \item  The $c$-number feature of  the null-$A$-mode field implies that it may take the form of
 an expectation value of some operator in the state $|0^{\rm np}\ra$.
  \item The above-mentioned operator describes emergence and vanishing of virtual $e$-$\ov e$-mode pairs in
  quantum fluctuations and, hence,  should contain both the $e$-$\ov e$-mode field $\psi(x)$
 and its conjugate field $\psi^\dag(x)$.
  \item To construct a vector field from $\psi(x)$ and $\psi^\dag(x)$, 
  the simplest method is to make use of $\gamma_\mu$ or $\pp_\mu$.
\end{enumerate}
 Since $\psi(x)$ and $\psi^\dag(x)$ do not act on the null $A$-mode state $|\infty^\text{np}\ra$,
 the null-$A$-mode field may in fact be written as an expectation value for the vacuum state $|0\ra$.

 Then, there are two simplest and most natural candidates for expression of $ A^{\rm np}_\mu(x)$.
 The first one is
\begin{gather}\label{epsi-1}
 A^{\rm np}_\mu(x) = \N_0( F^{(1)}_\mu + F^{(2)}_\mu),
\end{gather}
 where
\begin{subequations}\label{Fmu-12}
\begin{gather}\label{Fmu-1}
 F^{(1)}_\mu= \la 0|\psi^\dag (x) \gamma^0\gamma_\mu \psi(x)|0\ra, \hspace{1.1cm}
\\ F^{(2)}_\mu = -\la 0|\tr \Big( \gamma^0\gamma_\mu \psi(x) \psi^\dag (x) \Big)|0\ra; \label{Fmu-2}
\end{gather}
\end{subequations}
 and the second one  is
\begin{gather}\label{epsi-2}
 A^{\rm np}_\mu(x) =  z \N_0 (f^{(1)}_\mu + f^{(2)}_\mu),
\end{gather}
 where $z$ is a parameter and
\begin{subequations}\label{fmu-12}
\begin{gather}\label{}
 f^{(1)}_\mu= \la 0|\psi^\dag (x) \gamma^0 \big( \pp_\mu \psi(x) \big)|0\ra, \hspace{1cm}
\\ f^{(2)}_\mu = - \la 0|\tr \Big( \gamma^0 (\pp_\mu \psi(x)) \psi^\dag (x)  \Big)|0\ra.
\end{gather}
\end{subequations}
 As shown in Appendix \ref{app-F=f}, under the plane-wave expansion of the $e$-$\ov e$-mode field,
 the two expressions of $A^{\rm np}_\mu(x)$ in Eq.(\ref{epsi-1})
 and Eq.(\ref{epsi-2}) are equivalent with $z=i/m_0$.

 Some remarks for $F^{(1,2)}_\mu$ (similar for $f^{(1,2)}_\mu$):
 (i) The term $F^{(1)}_\mu$ in fact describes an effect  of emergence and vanishing of virtual $\ov e$-mode,
 while, the term $F^{(2)}_\mu$ is for virtual $e$-mode.
 (ii) The minus sign on the rhs of Eq.(\ref{Fmu-2})
 is due to the exchange of the order of $\psi$ and $\psi^\dag$, compared with that on the rhs of Eq.(\ref{Fmu-1}).

 Substituting  Eq.(\ref{psi-psi+-QED}) into Eq.(\ref{fmu-12}) and making use of Eq.(\ref{<Ur|Us>-sp}),
 one finds that
\begin{subequations}\label{fmu-12-fin}
\begin{gather}\label{fmu-12-fin-1}
 f^{(1)}_\mu = i \int d\ww p V^{r\dag }  (\bp) \gamma^0 p_\mu V^{r}  (\bp)  = -4im_0 \int d\ww p p_\mu ,
 \\  f^{(2)}_\mu = i \int d\ww q U^{r\dag }  (\bq) \gamma^0 q_\mu U^{r}  (\bq) = 4im_0 \int d\ww q  q_\mu. \label{fmu-12-fin-2}
\end{gather}
\end{subequations}
 To deal with the integrals on the rhs of the two subequations of Eq.(\ref{fmu-12-fin}), 
 one may employ a momentum regularization scheme.
 In this scheme, one considers a finite three-momentum region with a cutoff $\Lambda$,
 i.e, with $|\bp| < \Lambda$ \cite{Gu13},
 denoted by $\Omega(\Lambda)$.
 Under this regularization, $f^{(1)}_\mu$ and $f^{(2)}_\mu $ are written as
\begin{subequations}\label{fmu-12-fin-L}
\begin{gather}\label{fmu-12-fin-1-L}
 f^{(1)}_{\mu, \Lambda} = -4im_0 \int_{\Omega(\Lambda)} d\ww p p_\mu ,
 \\  f^{(2)}_{\mu, \Lambda} =  4im_0 \int_{\Omega(\Lambda)} d\ww q  q_\mu. \label{fmu-12-fin-2-L}
\end{gather}
\end{subequations}
 Clearly, $f^{(1)}_{\mu, \Lambda} + f^{(2)}_{\mu, \Lambda} =0$ and, as a result,
 one gets that $f^{(1)}_\mu + f^{(2)}_\mu =0$ in the limit of $\Lambda \to \infty$.
 Hence, 
\begin{gather}\label{Anp=0}
 A^{\rm np}_\mu(x)=0.
\end{gather}
 (See Appendix \ref{app-f1+f2=0} for a justification of Eq.(\ref{Anp=0})
 in view of the dynamics of virtual processes.)

\subsection{Null-$A$-mode field under gauge transformation}\label{sect-gauge-null-A}

 As seen in Eq.(\ref{Anp=0}),  the null-$A$-mode field vanishes
 under the plane-wave expansion of the $e$-$\ov e$-mode field given in Eq.(\ref{psi-psi+-QED}).
 In this section, we show that this field gets finite values under the gauge transformations of the $e$-$\ov e$-mode field 
 given in Eq.(\ref{psi-e-gt-2}).

 With the null-$A$-mode field included, the total $A$-mode field, denoted by $A^{\rm tot}_{\mu}(x)$, is written as
\begin{gather}\label{Acon-two}
 A^{\rm tot}_{\mu}(x) = A_\mu(x) + A^{\rm np}_\mu(x).
\end{gather}
 Correspondingly, the total Lagrangian density $\LL^{\rm tot}_{\rm QED}$ is written as
\begin{gather}\label{L0-ep-tot}
 \LL^{\rm tot}_{\rm QED} = \psi^\dag \gamma^0 (i \gamma^\mu D^{\rm tot}_\mu -m_0)\psi 
 - \frac{1}{4} (F^{\rm tot}_{\mu\nu})^2,
\end{gather}
 where 
\begin{gather}\label{Dmu-QED-tot}
 D^{\rm tot}_\mu = \pp_\mu - i eA^{\rm tot}_{\mu}(x).
 \\ F^{\rm tot}_{\mu\nu} = \pp_\mu A^{\rm tot}_\nu - \pp_\nu A^{\rm tot}_\mu.
\end{gather}

 In the previous section, two expressions of $A^{\rm np}_\mu(x)$ were given 
 [Eq.(\ref{epsi-1}) and Eq.(\ref{epsi-2})],
 which are identical under the plane-wave expansion of the $e$-$\ov e$-field. 
 It is easy to see that these two expressions give different predictions 
 under the local gauge transformation in Eq.(\ref{psi-e-gt-2}).
 In fact, the rhs of Eq.(\ref{Fmu-12}) does not change under the transformation
 and, hence, the expression in Eq.(\ref{epsi-1}) predicts no change of the null-$A$-mode field;
 in other words, it always predicts a vanishing null-$A$-mode field. 
 In contrast, Eq.(\ref{epsi-2}) predicts that the field may get a finite value.

 To see which of the two expressions  of $A^{\rm np}_\mu(x)$ is appropriate 
 under the local gauge transformations in Eq.(\ref{psi-e-gt-2}), let us consider a special case of the 
 gauge transformation, given by $\theta(x) = i q x$ with a constant vector $q^\mu$.
 Moving the gauge-phase terms of $e^{-iqx} $ and $e^{iqx}$ into the integrations for the quantum fields of 
 $\ww \psi(x)$ and $\ww \psi^\dag(x)$, respectively, it is seen that
 the creation operator part for the $e$-mode is written as $b^{r\dag}(\bp) U^{\dag r}(\bp) e^{i(p+q)x}$, 
 while,  the corresponding part for the $\ov e$-mode is written as $d^{r\dag}(\bp) V^{r}(\bp) e^{i(p-q)x}$
 with a minus sign before the phase term $iqx$.
 Thus, this specific gauge transformation may be interpreted as causing momentum shift for the
 $e$-mode and $\ov e$-mode in the corresponding parts of their fields,
 and the shift is different for these two modes.

 We recall that, in the plane-wave-expansion case discussed in the previous section,
 the null-$A$-mode field vanishes, meanwhile, the interaction Lagrangian $\LL_{\rm int}(x)$ predicts that
 the $e$-mode and $\ov e$-mode generated from a null $A$-mode should possess opposite momenta. 
 These two facts, together with the above-discussed interpretation of that specific gauge transformation, 
 suggests that the null-$A$-mode should not definitely vanish under the gauge transformation.

 Based on discussions given above, we assume that Eq.(\ref{epsi-2}) is appropriate
 in the computation of the null-$A$-mode field under gauge transformations.
 Substituting Eq.(\ref{psi-e-gt-2}) into Eq.(\ref{fmu-12}), straightforward derivation shows that
\begin{subequations}\label{fmu-12-gt}
\begin{gather}\label{fmu-12-gt-1}
 f^{(1)}_\mu \to \ww f^{(1)}_\mu = f^{(1)}_\mu  +4m_0i (\pp_\mu \theta )\int d\ww p ,
 \\  f^{(2)}_\mu \to \ww f^{(2)}_\mu = f^{(2)}_\mu  +4m_0i (\pp_\mu \theta )\int d\ww p . \label{fmu-12-gt-2}
\end{gather}
\end{subequations}
 This gives that
\begin{gather}\label{ww-Anp}
 A^{\rm np}_\mu(x)  \to  \ww A^{\rm np}_\mu(x) =  -  8\N_0  \big(\pp^\mu \theta(x) \big)\int d\ww p.
\end{gather}
 Under the momentum regularization discussed previously, the above transformation is written as
\begin{gather}\label{ww-Anp-L}
 A^{\rm np}_{\mu, \Lambda}(x)  \to  \ww A^{\rm np}_{\mu, \Lambda}(x) 
 =  -  8\N_0  \big(\pp^\mu \theta(x) \big)\int_{\Omega(\Lambda)} d\ww p.
\end{gather}
 We set the normalization factor $\N_0$ as
\begin{gather}\label{N0}
 \N_0 = \left( 8 e \int_{\Omega(\Lambda)} d\ww p \right)^{-1}.
\end{gather}
 Then, in the limit of $\Lambda \to \infty$, we get  the following expression of 
 the transformed null-$A$-mode field, 
\begin{gather}\label{Anp-gt}
 \ww A^{\rm np}_\mu(x) = -\frac 1e  \pp_\mu \theta(x).
\end{gather}

 Equation (\ref{Anp-gt}) shows that the transformed null-$A$-mode field $\ww A^{\rm np}_\mu(x)$ 
 is equal to what is usually regarded as the gauge-symmetry-required change of the field $A_\mu(x)$
 [see Eq.(\ref{Amu-gt-QED})].
 Since now $A^{\rm np}_\mu(x)$ and $A_\mu(x)$ form the total field $A^{\rm tot}_\mu(x)$, 
 in order to keep the total Lagrangian $\LL^{\rm tot}_{\rm QED}$ in Eq.(\ref{L0-ep-tot})
 invariant under the gauge transformations of the $e$-$\ov e$-mode field in Eq.(\ref{psi-e-gt-2}), 
 one needs to keep the field $A_\mu(x)$ unchanged, that is, to assume that 
\begin{gather}\label{}
 \ww A_\mu(x) = A_\mu(x).
\end{gather}

 Thus, with the  null-$A$-mode field $ A^{\rm np}_\mu(x)$ included,
 one gets a natural explanation to the gauge-symmetry requirement that the total gauge field should change 
 by the term $-\frac 1e  \pp_\mu \theta(x)$, 
 accompanying the gauge transformation of the $e$-$\ov e$-mode field in Eq.(\ref{psi-e-gt-2}).
 In other words, the total Lagrangian written in the form of Eq.(\ref{L0-ep-tot})
 naturally possesses the local $U(1)$ gauge symmetry under the transformation in Eq.(\ref{psi-e-gt-2}).

\section{ Summary and discussions}\label{sect-conclusion}

 In this paper, we go further along a direction of line discussed in Ref.\cite{Weinberg-book},
 for the purpose of formulating QED completely on a quantum ground.
 A key point of our approach is to move away an usually-adopted assumption from the fundamental level
 of the theory,
 which basically states that the spin spaces of modes (bare particles) should possess properties similar to 
 those of the related free particles observed experimentally. 
 With this assumption released,
 a quite simple formulation of QED is found at  the fundamental level (without resorting to any classical field),
 which is done by three steps:  (i) establishing the quantum state space,
 (ii) constructing the quantum fields, and (iii) building the total Lagrangian by gauge symmetry.

 Within this formulation of QED,   the photonic field includes,
 in addition to the ordinarily-discussed part, 
 a null-$A$-mode part that is related to $A$-modes (bare photons) with zero momentum. 
 An explicit expression of this part of the photonic  field is derived, 
 which reflects a mean effect of quantum fluctuations in the vacuum.
 It predicts that, under local $U(1)$ gauge transformations of the fermionic field, 
 the null-$A$-mode part of the bosonic field undergoes a change, which has the same form as 
 the well-known gauge-symmetry-required change of the gauge field. 
 This suggests that the change of the photonic  field, which is required by  gauge transformations, should
 come from its null-$A$-mode part.
 With this understanding, the total Lagrangian naturally keeps invariant 
 under local $U(1)$ gauge transformations of the fermionic field.

 The above-discussed mechanism, by which the photonic  field changes under gauge transformations, 
 is neglected in the ordinary formulations of QED.
 This is because there the photonic  field is introduced by quantizing the classical electromagnetic field;
 while, as is well known, the component of a classical electromagnetic (vector) field,
 which corresponds to zero frequency and zero wave length, has no physical significance.
 \footnote{ In fact, a hint may be seen even in the ordinary formulations of QED.
 The hint is given by the fact that the  bare-photon state with zero momentum 
 corresponds a singular point in the momentum space,
 since the integration over its neighborhood
 may give divergent results for Feynman diagrams (infrared divergence).}

 In this paper,  we use the term ``mode'' to refer to what is usually called bare elementary particle.
 One reason of using this name was given previously, i.e., to  
 make an explicit distinction from the concept of free particle observed experimentally. 
 Here, we give another reason, which is that the term mode may be assigned a more generic meaning.

 In fact, when quantum states of modes are introduced as the first step in the formulation of QED, 
 they are labelled with momentum and spin indexes;
 in other words,  no definite feature in the configuration space is assigned to them.
 It is quantum fields, which are constructed at the second step from wave functions of modes
 together with creation-annihilation operators,
 that may be related to some specific feature(s) in the configuration space. 
 For example, through a plane-wave expansion, one may get some point feature. 
 In principle, exploiting mode wave functions other than plane waves, it is possible to 
 construct quantum fields with other features (say, a string feature).

 In future investigation, it should be of interest to study whether  the method used here
 may be generalized to the electroweak theory with the gauge symmetry $U(1)\otimes SU(2)$.
 When doing this,  one may consider a stage of the theory
 before the Higgs mechanism is used to introduce masses to vector bosons;
 at this stage, there are two species of massless neutral  boson.
 Since the $SU(2)$ gauge symmetry involves an intrinsic degree of freedom,
 such a generalization (if applicable) can not be a straightforward one and some modification is expected.

 \acknowledgements

 The author is grateful to Yan Gu and Hong Zhao for valuable discussions and suggestions.
 This work was partially supported by the National Natural Science Foundation of China under Grant
 Nos.~11535011 and 11775210.

\appendix

\section{Relation between Eqs.(\ref{epsi-1}) and (\ref{epsi-2})}\label{app-F=f}

 In this appendix, we show that the two expressions in Eqs.(\ref{epsi-1}) and (\ref{epsi-2}) are equivalent with $z=i/m_0$.
 Substituting Eq.(\ref{psi-psi+-QED}) into Eq.(\ref{Fmu-12}), one finds that
\begin{gather}\notag
  F^{(1)}_\mu = \la 0|    \int  d\ww q \left(  b^{s\dag}  (\bq)
 U^{s\dag }  (\bq) e^{iqx} +d^s(\bq) V^{s\dag }  (\bq)e^{-iqx} \right)
 \\ \gamma^0 \gamma_\mu  \int d\ww p \left(  b^r  (\bp) U^{r}  (\bp) e^{-ipx}
  + d^{r\dag}(\bp) V^{r}  (\bp) e^{ipx} \right)|0\ra \notag
 \\  = \la 0|    \int d\ww p d\ww q d^s(\bq) V^{s\dag }  (\bq)e^{-iqx}
 \gamma^0 \gamma_\mu   d^{r\dag}(\bp) V^{r}  (\bp) e^{ipx} |0\ra \notag
 \\  = \int d\ww p V^{r\dag }  (\bp) \gamma^0 \gamma_\mu V^{r}  (\bp).
\end{gather}
 Similarly,
\begin{gather}\notag
 F^{(2)}_\mu = -\la 0| \tr \Big[ \gamma^0 \gamma_\mu
  \int d\ww p \big(  b^r  (\bp) U^{r}  (\bp) e^{-ipx}
  + d^{r\dag}(\bp) V^{r}(\bp)
  \\ \times e^{ipx} \big)
   \times \int  d\ww q \left(  b^{s\dag}  (\bq)
 U^{s\dag }  (\bq) e^{iqx} +d^s(\bq) V^{s\dag }  (\bq)e^{-iqx} \right) \Big]|0\ra\notag
 \\  = -\int d\ww p \ \tr \Big[  \gamma^0 \gamma_\mu U^{r } (\bp) U^{r \dag}  (\bp) \Big]\notag
  \\ = -\int d\ww p \  U^{r \dag}  (\bp)  \gamma^0 \gamma_\mu U^{r } (\bp)
\end{gather}

 The Dirac equation $(\gamma^\nu p_\nu +m_0) V^r(\bp)=0$ gives that
\begin{subequations}\label{V-V+}
\begin{gather}\label{stat-DE-V}
  V^r(\bp) = -   \gamma^\nu p_\nu  V^r(\bp) / m_0,
 \\   V^{r\dag}(\bp) = - V^{r\dag} \gamma^{\nu\dag} p_\nu /m_0. \label{stat-DE-Vdag}
\end{gather}
\end{subequations}
 Making use of Eq.(\ref{V-V+}) and the relations that $(\gamma^0)^2=1$ and
 $\gamma^0 \gamma^{\mu\dag}  \gamma^0 = \gamma^\mu$, one finds that
\begin{subequations}\label{}
\begin{gather}\label{}
   V^{r\dag }  (\bp) \gamma^0 \gamma^\mu V^{r}  (\bp)
  = - V^{r\dag }   \gamma^0 \gamma^\mu  \gamma^\nu p_\nu  V^{r}/m_0,
 \\   V^{r\dag }  (\bp) \gamma^0 \gamma^\mu V^{r}  (\bp)
  = - V^{r\dag } \gamma^0 \gamma^{\nu} p_\nu    \gamma^\mu V^{r} /m_0.
\end{gather}
\end{subequations}
 Then, noting the relation $\{ \gamma^\mu,  \gamma^\nu \} =2g^{\mu\nu}$ and Eq.(\ref{<Ur|Us>-sp}), one gets that
\begin{gather} \label{VggV=pmu}
  V^{r \dag }  (\bp) \gamma^0 \gamma^\mu V^{r}  (\bp)
 = - \frac 1m_0 V^{r\dag }   \gamma^0 g^{\mu\nu}   p_\nu   V^{r} = 4   p^\mu.
\end{gather}

 Following the same procedure as that given above, for the spinors $U^r(\bp)$,
 which satisfy the Dirac equation $ (\gamma^\mu p_\mu -m_0) U^r(\bp)=0$,
 one finds that
\begin{gather}\label{UggU=pmu}
 U^{r\dag }  (\bp) \gamma^0 \gamma^\mu U^{r}  (\bp) =  4  p^\mu.
\end{gather}
 Finally, making use of Eq.(\ref{fmu-12-fin}) and Eqs.(\ref{VggV=pmu})-(\ref{UggU=pmu}), one gets
\begin{gather}\label{}
 f^{(1,2)}_\mu = -(im_0) F^{(1,2)}_\mu
\end{gather}
 and this accomplishes the proof.

\section{Another justification of Eq.(\ref{Anp=0})}\label{app-f1+f2=0}

 In this appendix, we give another justification for Eq.(\ref{Anp=0}),
 which is based on the dynamics of virtual processes. 
 To this end, we note a fact mentioned previously,
 that is, the term $f^{(1)}_\mu$ is for emergence and vanishing of virtual $\ov e$-modes,
 while, $f^{(2)}_\mu$ is for virtual $e$-modes.
 Since the $e$-mode and $\ov e$-mode in a virtual pair generated by a null $A$-mode have opposite momenta,
 a point $\bp$ on the rhs of Eq.(\ref{fmu-12-fin-1})
 should correspond to a point $\bq = -\bp$ on the rhs of Eq.(\ref{fmu-12-fin-2}).
 This gives that $f^{(1)}_0 +  f^{(2)}_0 =  0$ and
\begin{gather}\label{fmu-1+2}
 f^{(1)}_i +  f^{(2)}_i =  8im_0 \int d\ww q  q_i, \quad i=1,2,3.
\end{gather}
 To give further evaluation, we note the physical picture
 that virtual pairs emerge randomly, such that there is no difference 
 between the probability for $\bq$ and that for $-\bq$.
 This implies that the contribution of a value of $q_i$ on the rhs of Eq.(\ref{fmu-1+2}) be canceled by that of $-q_i$,
 resulting in a vanishing result of the integration.



\begin{thebibliography}{99}

 \bibitem{Weinberg-book} S.~Weinberg, {\it The Quantum Theory of
 Fields} (Cambridge University Press, Now York, 1996).
 \bibitem{Peskin} M.E. Peskin and D.V. Schroeder, {\it In Introduction to Quantum Field Theory}
 (Westview Press, 1995).
 \bibitem{Itzy} C.~Itzykson and J.~B.~Zuber,
 {\it Quantum Field Theory} (McGraw-Hill, New York, 1980).


\bibitem{little-group} G.~W.~Mackey, Ann.~Math.~{\bf 55}, 101 (1952); {\bf 58}, 193 (1953);
 Acta.~Math.~{\bf 99}, 265 (1958);
 {\it Induced Representations of Groups and Quantum Mechanics} (Benjamin, New York, 1968). 

 \bibitem{Pauli40} W.~Pauli, Phys.Rev.~{\bf 58}, 716 (1940).
 \bibitem{SW64} R.F.~Streater and A.S.~Wightman, {\it PCT, Spin and Statistics, and
 All That} (Benjamin/Cummings, Reading, Mass., 1964).


\bibitem{pra16-commu} W.-g. Wang, Phys.Rev.A {\bf 94}, 012112 (2016).


\bibitem{Gupta} S.N. Gupta, Proceedings of Physical Society A {\bf 63}, 681 (1950).

\bibitem{Gu13} As shown in a paper of Y.~Gu, Phys.~Rev.~A {\bf 88}, 042103 (2013),  under this momentum
regularization scheme, QED can be formulated in a gauge-covariant way.

\end{thebibliography}
\end{document}